\documentclass[twocolumn,aps,pra,superscriptaddress,10pt]{revtex4-2}
\usepackage{graphicx}
\usepackage{amsmath}
\usepackage{amssymb}
\usepackage{braket}
\usepackage{bm}
\usepackage{dcolumn}
\usepackage{subfigure}
\usepackage{graphicx}
\usepackage[colorlinks=true,linkcolor=blue,citecolor=blue,urlcolor=blue]{hyperref}
\usepackage{soul}
\usepackage{changes}
\usepackage{amsthm,amsmath,amssymb}

\usepackage{mathrsfs}
\usepackage{multirow}

\makeatother

\begin{document}
%\begin{CJK*}{GB}{} 
%,twoside

\title{Bayesian Interpretation of Husimi Function and  Wehrl Entropy}%in an atomic heteronuclear spinor Bose gas}%of Ultracold Spin Mixture }

\author{Chen Xu}
%\email{Present address: Homer L. Dodge Department of Physics and Astronomy, The University of Oklahoma, 440 W. Brooks Street, Norman, Oklahoma 73019, USA}

\affiliation{Department of Physics, Renmin University of China, Beijing, 100872,
China}

\author{Yiqi Yu}
%\email{Present address: Homer L. Dodge Department of Physics and Astronomy, The University of Oklahoma, 440 W. Brooks Street, Norman, Oklahoma 73019, USA}

\affiliation{Department of Physics, Renmin University of China, Beijing, 100872,
China}
%\author{Dazhi Xu}
%\email{dzxu@bit.edu.cn}
%
%\affiliation{Center for Quantum Technology Research and School of Physics, Beijing
%Institute of Technology, Beijing 100081, China}

\author{Peng Zhang}
\email{pengzhang@ruc.edu.cn}

\affiliation{Department of Physics, Renmin University of China, Beijing, 100872,
China}

\affiliation{Key Laboratory of Quantum State Construction and Manipulation (Ministry of Education), Renmin University of China, Beijing, 100872, China}

\affiliation{Beijing Key Laboratory of Opto-electronic Functional Materials \&
Micro-nano Devices, Renmin University of China, Beijing, 100872, China}

\date{today}

\begin{abstract}
Husimi function (Q-function) of a quantum state is the distribution function of the density operator in the coherent state representation. It is widely used in theoretical research, such as in quantum optics. 
The Wehrl entropy is the Shannon  entropy of the Husimi function, and is non-zero even for pure states. This entropy
 has been extensively studied in mathematical physics. Recent research also suggests a significant connection between the Wehrl entropy and many-body quantum entanglement in spin systems.
We investigate the statistical interpretation of the Husimi function and the Wehrl entropy, taking the system of $N$ spin-1/2 particles as an example.
Due to the completeness of coherent states, the Husimi function and Wehrl entropy can be explained 
via the positive operator-valued measurement (POVM) theory, although the coherent states are not a set of orthonormal basis.
Here, with the help of the Bayes' theorem, 
we provide an alternative probabilistic interpretation for the Husimi function and the Wehrl entropy. This interpretation is based on direct measurements of the system, and thus does not require the introduction of an ancillary system as in POVM theory.
Moreover, under this interpretation 
the classical correspondences of the Husimi function and Wehrl entropy
are just phase-space probability distribution function of $N$ classical tops, and its associated entropy, respectively. Therefore, this explanation 
contributes to a better understanding of the relationship between the Husimi function, Wehrl entropy, and classical-quantum correspondence. The generalization of this statistical interpretation to continuous-variable systems is also discussed.

%The Wehrl entropy  of  a  quantum state is the entropy of the coherent-state distribution function (Husimi function) of this state. This entropy is non-zero even for pure states. We investigate  the Wehrl entropy for $N$ spin-1/2 particles with respect to  the SU(2)$^{\otimes N}$ coherent states ({\it i.e.}, the direct products of spin coherent states of each particle).
%We focus on the statistical interpretation of this Wehrl entropy.  Despite the coherent states not forming a group of orthonormal bases, 
%we prove that the Wehrl entropy can still be interpreted as the entropy of a  probability distribution with clear physical meaning. 

\end{abstract}
\maketitle
%\end{CJK*} 

\textbf{Keywords:} Bayesian interpretation, Husimi function, Wehrl entropy, classical-quantum correspondence

\section{Introduction}
The Husimi function (Q-function) \cite{Husimi1940} of a quantum state
described by the density operator  $\hat \rho$,
is the distribution function in the coherent state representation. 
For instance, the Husimi function of a harmonic oscillator
and a spin-1/2 particle can be expressed as
$\langle{{\bm\alpha}}|{{\hat \rho}}|{\bm\alpha}\rangle/\pi$ and $\langle{{\bm n}}|{{\hat \rho}}|{\bm n}\rangle/2\pi$, 
respectively. Here $\{|{\bm\alpha}\rangle\}$ are the coherent states of this harmonic oscillator, and $\{|{\bm n}\rangle\}$ are the SU(2) spin-coherent states. 
Husimi function has broad applications in various directions of quantum physics, {\it e.g.}, quantum optics and quantum precise measurement. 

The
Wehrl entropy
of a  quantum state
 $\hat \rho$
 is  the Shannon entropy of the corresponding Husimi function \cite{Wehrl1979}.
Wehrl entropy has been extensively studied in mathematical physics \cite{Lieb1978,Lee1988,Sugita2003,Lieb2014} and quantum chaos \cite{sugita2002second}. 
For many-body spin systems, it is also found that  this entropy
 can be used as a  measurement of the complexity of the many-body entanglement \cite{Sugita2003,xu2024wehrl}.

Here we investigate the statistical interpretation of Husimi function and Wehrl entropy. 
We take the system of $N$ spin-1/2 particles as an example.

Since the coherent states are not a group of orthonormal basis of the Hilbert space, the Husimi function cannot be directly interpreted as the probability distribution of the measurement outcomes of a certain observable. Thus, the Wehrl entropy cannot
be directly interpreted as the entropy corresponding to such a probability distribution, which distinguishes it from the more commonly used von Neumann entropy. However, 
due to the completeness relation of the coherent states, {\it i.e.},
$\int \mathrm{d}{\bm n}|{\bm n}\rangle\langle{{\bm n}}|/Z= \hat I$, with $Z$ being a constant, 
 the
projection operators $\{|{\bm n}\rangle\langle {\bm n}|\}$ of the coherent states
  form a positive operator-valued measurement (POVM). Explicitly \cite{povm}, one can 
implement this POVM 
({\it i.e.}, realize a measurement with the probability distribution of the possible results being same as the Husimi function)
 by first entangling the quantum systems with an ancillary system, and then measuring the ancillary system. Such realization of the POVM can be understood as a statistical interpretation of the Husimi function. For systems with continuous variables, such as photons, this implementation of POVM has been demonstrated in various experiments using heterodyne detection \cite{Weedbrook2012,Mueller2016}. 

In this work, using the  Bayes theorem we
  provide
an alternative probabilistic interpretation  for 
   the Husimi function and the Wehrl entropy, which has the following two characteristics:
\begin{itemize}
\item[(1)] 
Our interpretation is based on the direct measurement of the quantum system. The aforementioned ancillary system of the POVM is not required.

\item[(2)] More importantly, under this interpretation, 
the classical correspondence of
the Husimi function is exactly
the Liouville function (phase-space probability distribution function) of $N$ classical tops, and thus the one of the Wehrl entropy is the Gibbs entropy of these tops, respectively.

\end{itemize}

\noindent Therefore, our probabilistic interpretation  provides a more intuitive understanding of  Husimi function and Wehrl entropy. 
This statistical interpretation can be generalized to the continuous-variable systems.
As mentioned before, Husimi function and Wehrl entropy are widely used in 
the researches of various quantum systems. Thus, our results are helpful for understanding the 
properties of these systems, which are described by  Husimi function and Wehrl entropy. Moreover, these results  also  
enhance our comprehension of the relationship between  Husimi function, Wehrl entropy, and the classical-quantum correspondence. 
 
The remainder  of this paper is as follows. In Sec.~\ref{def}
we introduce the definitions of the Husimi function and Wehrl entropy of $N$ spin-1/2 particles.  The  statistical interpretation of the Wehrl entropy is given in Sec.~\ref{pma}. In Sec.~\ref{cc}, we describe the classical correspondence of the statistical interpretation of Husimi function and Wehrl entropy  of spin-1/2 particles. In Sec.~\ref{cv} we generalize the statistical interpretation to continuous-variable systems.
A summary of our results is given in Sec.~\ref{summary}, and some details of the calculation are given in the Appendix.

\section{The Husimi Function and Wehrl entropy of Spin Particles }
\label{def}

%{\color{red}The Husimi function of the Harmonic osillator or other systems in real space is defined as
%\begin{eqnarray}
	%P_H({\hat \rho};  {\bm\alpha})\equiv\frac{1}{(2\pi)^N}\langle{{\bm\alpha}}|{{\hat \rho}}|{\bm\alpha}\rangle,
%\end{eqnarray}
%where the complex number $\bm\alpha$ and the coherent state $|{\bm \alpha}\rangle$ satisfy $\hat a_j|{\bm \alpha}\rangle=\bm\alpha|{\bm \alpha}\rangle$, $\hat a_j$ is  the annihilation operator for the particle $j$.

%This is known as the Wehrl entropy of this system
%\begin{eqnarray}
%	S_W({\hat \rho})\equiv-\int d^{2N}{\bm \alpha}P_H({\hat \rho};  {\bm \alpha})\ln\bigg[P_H({\hat \rho};  {\bm \alpha})\bigg].
%\end{eqnarray}

%In phase space $(x,p)$,  the Wehrl entropy can be written as
%\begin{eqnarray}
%	S_W({\hat \rho})\equiv-\int \frac{d^{N}x d^{N}p}{\hbar^N} P_H({\hat \rho};  {\bm x,\bm p})\ln\bigg[P_H({\hat \rho};  {\bm x,\bm p})\bigg].
%\end{eqnarray}}

%%%%%%%%%%%%%%%%%%%%%%%%%%%%%%%%%%%%%%%%%%%%%%%%%%%%%%%%%%%%%%%%%%%%%%%%%%%%
We consider a system of $N$ spin-1/2 particles labled as $1,...,N$.  A spin coherent state  can be written as the direct product of the spin-coherent states for each individual particle:
\begin{eqnarray}
|{\bm {n}}\rangle\equiv|{{\bm n}}_1\rangle_1\otimes|{{\bm n}}_2\rangle_2\otimes...\otimes|{{\bm n}}_N\rangle_N,\label{sc}
\end{eqnarray}
with ${\bm n}\equiv({{\bm n}}_1,{{\bm n}}_2,...,{{\bm n}}_N)\in S^{2\otimes N},$
where ${{\bm n}}_j$ denotes a unit vector in three-dimensional space, corresponding to a point on the unit sphere (also known as the Bloch sphere), $S^2$, with $j=1, \ldots, N$. The unit vector can be expressed in spherical coordinates as ${\bm n}_j = (\sin\theta_j \cos\phi_j, \sin\theta_j \sin\phi_j, \cos\theta_j)$, where $\theta_j \in [0, \pi]$ and $\phi_j \in [0, 2\pi]$. In addition, $S^{2\otimes N}$ represents the Cartesian product of $N$ unit spheres.

Furthermore, $|{{\bm n}}_j\rangle_j$ is the spin-coherent state of the $j$th particle oriented along the direction ${{\bm n}}_j$, which have two eigenvalues $\pm 1$ and corresponding eigenstates,
\begin{eqnarray}
\left[{\hat {\bm \sigma}}^{(j)}\cdot  {{\bm n}}_j\right]|\pm{{\bm n}}_j\rangle_j=\pm|\pm{{\bm n}}_j\rangle_j,
\end{eqnarray}
where ${\hat{\bm \sigma}}^{(j)}=(\hat{\sigma}_x^{(j)}, \hat{\sigma}_y^{(j)}, \hat{\sigma}_z^{(j)})$, with $\hat{\sigma}_{x,y,z}^{(j)}$ being the Pauli operators associated with particle $j$  ($j=1,...,N$). Although the spin coherent states $\{|{\bm {n}}\rangle\}$ are not orthogonal, they satisfy the overcomplete relation
\begin{eqnarray}
\frac{1}{(2\pi)^N}\int \mathrm{d}{\bm n}|{\bm n}\rangle\langle{{\bm n}}|= \hat I,\label{com}
\end{eqnarray}
where ${\hat I}$ is the unit operator, and the integral of $\bm n$ can be written as
\begin{eqnarray}
\int \mathrm{d}{\bm n}=\prod_{j=1}^N\int_0^{2\pi}\mathrm{d}\phi_j\int_0^\pi\sin\theta_j\mathrm{d}\theta_j.
\end{eqnarray}

The Husimi function $P_H({\hat \rho}; {\bm n})$ of this system is defined as
\begin{eqnarray}
P_H({\hat \rho};  {\bm n})\equiv\frac{1}{(2\pi)^N}\langle{{\bm n}}|{{\hat \rho}}|{\bm n}\rangle,\label{hf}
\end{eqnarray}
where ${\hat \rho}$ is the density matrix representing the quantum state of the $N$ particles. Unlike the Wigner W-distribution, the Husimi function is both non-negative and bounded,
which follows that $0\leq P_H({\hat \rho}; {\bm n}) \leq 1/(2\pi)$ for any ${\bm n}$. Moreover,
the Husimi function $P_H({\hat \rho}; {\bm n})$ is normalized to unity
\begin{eqnarray}
\int \mathrm{d}{\bm n}P_H({\hat \rho};  {\bm n})=1.
\end{eqnarray}
Each Husimi function $P_H({\hat \rho}; {\bm n})$ uniquely corresponds to an $N$-particle quantum state ${\hat \rho}$. Specifically, if ${\hat \rho} \neq {\hat \rho}^\prime$, there must exist a direction ${\bm n} \in S^{2\otimes N}$ such that $P_H({\hat \rho}; {\bm n}) \neq P_H({\hat \rho}^\prime; {\bm n})$.

The Wehrl entropy for this system of $N$ spin-1/2 particles is defined as the entropy associated with the Husimi function:
\begin{eqnarray}
S_W({\hat \rho})\equiv-\int P_H({\hat \rho};  {\bm n})\ln\bigg[P_H({\hat \rho};  {\bm n})\bigg]\mathrm{d}{\bm n}.\label{sw}
\end{eqnarray}
Obviously, the Wehrl entropy is a functional of the quantum state ${\hat \rho}$, and it is in the form of the Shannon entropy of the Husimi function.

%It is clear that  this  WE 
%$S_W(|\psi\rangle\langle\psi|)$  is non-zero even  when the system is in a pure state.

\section{Statistical Interpretation of the Husimi Function and the Wehrl Entropy}
\label{pma}

In this section we demonstrate the statistical interpretation of the Husimi function and the Wehrl entropy. 

Since 
the   coherent states
$|{\bm n}\rangle$ and $|{\bm n}'\rangle$ defined above are not orthogonal if $\langle {\bm n}'_j|{\bm n}_j \rangle\neq 0$ for  all $j=1,...,N$, these  states are not eigen-states of the same observable.
Consequently, the Husimi function $P_H({\hat \rho};  {\bm n})$ cannot be interpreted as  the
probability distribution of the outcome of a measurement for any specific observable. Hence, the Wehrl entropy 
cannot be directly interpreted as the entropy of such a probability distribution.

Here we provide a statistical interoperation  for the Husimi function and the Wehrl entropy. We will demonstrate that
 they can still
be interpreted as  a certain well-defined  probability distribution with a clear physical meaning, and the corresponding entropy, respectively. 
%In addition, 
%in the classical analogy of our system ({\it i.e.}, when our system is analogized as $N$ classical rotators), 
%this probability distribution returns to the phase-space distribution, and thus the Wehrl entropy becomes to the Gibbs entropy. 

\begin{figure}[tbp]
 	\centering
 	\includegraphics[width=8.5cm]{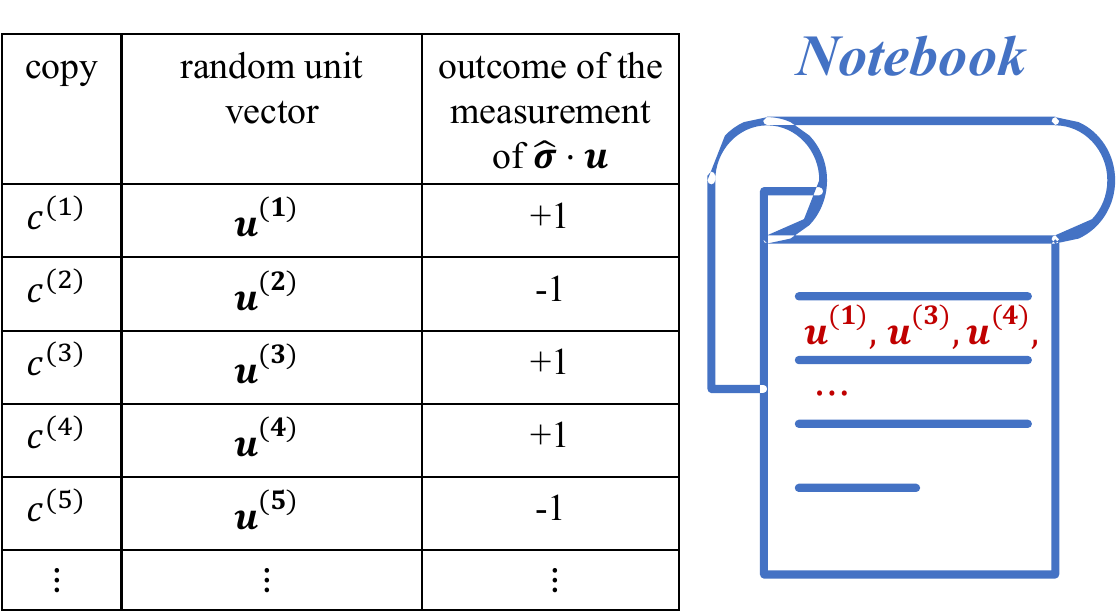}
 	\caption{A schematic  illustration of the though experiment of Sec.~\ref{pma} for a single spin-1/2 particle. As shown in the table, in the  $\eta$-th step ($\eta=1,2,..,m$), the experimenter measures the observable
 		${\hat {\bm \sigma}}\cdot {\bm u}^{(\eta)}$ 
 		for the copy $\eta$. If and only if the outcome is +1, then the experimenter notes down the direction  ${\bm u}^{(\eta)}$ in the notebook. In the case of this figure the outcome of the measurements 1, 3, 4,... are +1, and thus the directions ${\bm u}^{(1,3,4,..)}$ are noted down in the notebook.}
 	\label{texp}
 \end{figure}

We begin from the single-particle case ($N=1$). Consider the following thought experiment: Assume  there are $m$ copies $c^{(1)}, c^{(2)},...,c^{(m)}$ of the spin-1/2 particle, with each one being in the same state ${\hat \rho}$. In addition, the experimenter generates $m$ independent random unit vectors ${\bm u}^{(1)}, {\bm u}^{(2)},...,{\bm u}^{(m)}$ from the unit sphere $S^2$. Explicitly,  
the probability densities of these vectors in $S^2$ are all $1/(4\pi)$, and are independent of each other. Then the experimenter performs the following $m$ steps:
In the $\eta$-th  step ($\eta=1,...,m$), the experimenter measures
the Pauli operator vector ${\hat{\bm \sigma}}$ along the direction ${\bm u}^{(\eta)}$ ({\it i.e.}, the observable ${\hat {\bm \sigma}}\cdot {\bm u}^{(\eta)}$)
for the  copy $c^{(\eta)}$. If and only if the outcome of this  measurement  is $+1$,  the experimenter notes down the direction ${\bm u}^{(\eta)}$ in a notebook. In Fig.~\ref{texp} we schematically  illustrate these $m$ steps.

 We   consider the cases where $m$ is very large, and assume that when all these $m$ steps are finished, there are $D$ directions noted down in the notebook. Then, the following question arises:

%The experimenter prepare a notebook, and then
%perform the following $m$ steps:
%\begin{itemize}[itemindent=0.7em]
%\item[{\it Step 1}:]Randomly choose a direction ${\hat{\bm n}}_{c1}$ from the sphere surface $S_2$, and measure the observable  ${\bm \sigma}\cdot {\hat{\bm n}}_{c1}$
%for the copy $c_1$. If the outcome is $+1$, then note the direction ${\hat{\bm n}}_{c1}$ in the notebook. If the the outcome is $-1$, do nothing.
%
%\item[{\it Step 2}:]Similar as above, randomly choose a direction ${\hat{\bm n}}_{c2}$, and measure ${\bm \sigma}\cdot {\hat{\bm n}}_{c2}$
%for the copy $c_2$. If the outcome is $+1$, note  ${\hat{\bm n}}_{c2}$ in the notebook. If the the outcome is $-1$, do nothing.
%\item[]...
%%\item[{\it Step m}:] Randomly choose a direction ${\hat{\bm n}}_{cm}$, and measure ${\bm \sigma}\cdot {\hat{\bm n}}_{cm}$
%%for the copy $c_m$. If the outcome is $+1$, note  ${\hat{\bm n}}_{cm}$ in the notebook. If the the outcome is $-1$, do nothing.
% \end{itemize}
% Namely, in each step the experimenter randomly choose a direction, and measure the ${\bm \sigma}$-operator along that direction for an individual copy, and denote this direction only if the outcome of the measurement is $+1$.
  
{\bf Question}: {What is the distribution of the  directions noted down in the notebook? In another word, for a given direction ${\bm n}\in S^2$,  how many  directions, among the $D$ ones in the notebook, are in a small solid angle $\Delta\Omega$ around ${\bm n}$?}

This question can be answered as follows:
  
{\bf Answer}:  Denote the answer to the above question ({\it i.e.}, the number of directions in the region described above) as ${\cal N}({\bm n})$.
Since a direction can be noted down in the notebook if and only if the outcome of the corresponding measurement is $+1$, the answer to the above question can be expressed as:
\begin{eqnarray}
{\cal N}({\bm n})=D{\cal P}\left({\bm n}\big\vert+1\right)\Delta\Omega.\label{nn1}
\end{eqnarray}
Here ${\cal P}\left({\bm n}\big\vert+1\right)$ is a {\it conditional probability density}, {\it i.e.}, the probability density of a randomly-chosen direction being ${\bm n}$, under the condition that the outcome of the measurement of the ${\hat {\bm \sigma}}$-operator along that direction is $+1$.

Moreover, according to 
Bayes' theorem, we have
 \begin{eqnarray}
{\cal P}\left({\bm n}\big\vert+1\right)=\frac{{\cal P}\left(+1\big\vert {\bm n}\right)}{{\cal P}\left(+1\right)}{\cal P}\left({\bm n}\right).\label{bp1}
\end{eqnarray}
Here
${\cal P}\left({\bm n}\right)=1/(4\pi)$ is the probability density of a randomly-chosen direction being ${\bm n}$,
 ${\cal P}\left(+1\vert{\bm n}\right)$ is the probability that outcome of the measurement of the ${\bm \sigma}$-operator along this random-chosen direction is $+1$ and can be expressed as 
  \begin{eqnarray}
{\cal P}\left(+1\big\vert {\bm n}\right)=\langle{{\bm n}}|{{\hat \rho}}|{\bm n}\rangle,\label{p1n}
\end{eqnarray}
and  ${\cal P}(+1)=\int \mathrm{d}{\bm n}^\prime
{\cal P}(+1\big\vert {\bm n}^\prime){\cal P}({\bm n}^\prime).
$
% is the probability that the outcome of the measurement of  ${\hat {\bm \sigma}}$ along this randomly-chosen direction is $+1$. 

Since both ${\cal P}\left({\bm n}\right)$ and ${\cal P}\left(+1\right)$ are independent of ${\bm n}$, Eqs.~(\ref{bp1}, \ref{p1n}) yield that 
$
{\cal P}\left({\bm n}\big\vert+1\right)=Z\langle{{\bm n}}|{{\hat \rho}}|{\bm n}\rangle,
$
with the constant $Z$ being determined by the normalization condition $\int \mathrm{d} {\bm n}{\cal P}\left({\bm n}\big\vert+1\right)=1$. Direct calculation gives $Z=1/(2\pi)$. Therefore, we finally obtain ${\cal P}\left({\bm n}\big\vert+1\right)=\langle{{\bm n}}|{{\hat \rho}}|{\bm n}\rangle/(2\pi)$, or
 \begin{eqnarray}
{\cal P}\left({\bm n}\big\vert+1\right)=P_H({\hat \rho};  {\bm n}).\label{pph}
\end{eqnarray}
Substituting this result into Eq.~(\ref{nn1}), we find that the amount of  the noted-down directions  in the small solid angle $\Delta\Omega$ among ${\bm n}$ is:
\begin{eqnarray}
{\cal N}({\bm n})=DP_H({\hat \rho};  {\bm n})\Delta\Omega.\label{nn}
\end{eqnarray}
That is the answer to the question. 

  \begin{figure}[tbp]
	\centering
	\includegraphics[width=8.3cm]{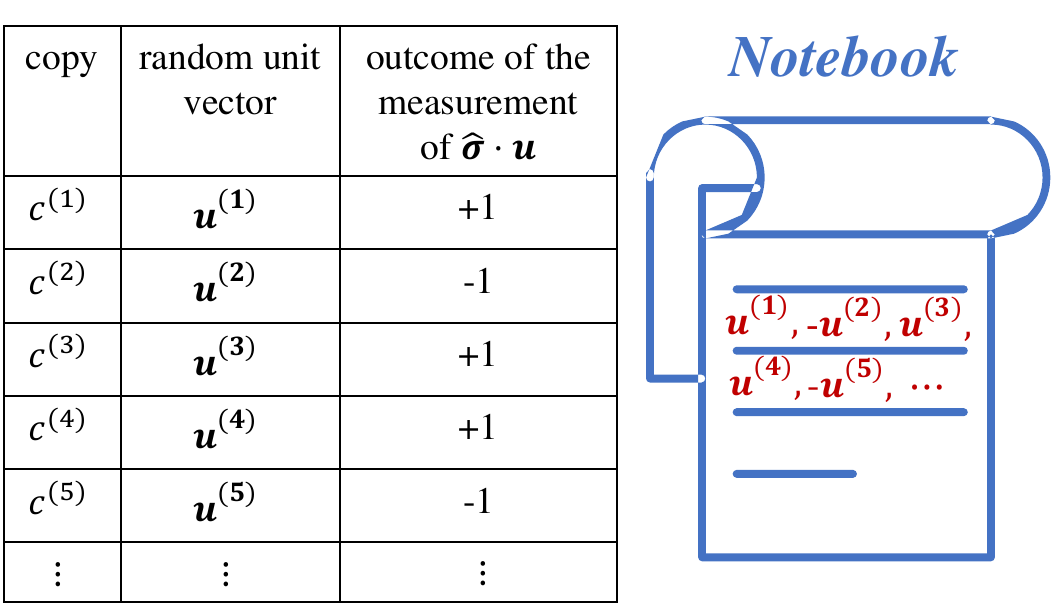}
	\caption{A schematic  illustration of the second though experiment of Sec.~\ref{pma} for a single spin-1/2 particle. As shown in the table, in the  $\eta$-th step ($\eta=1,2,..,m$), the experimenter measures the observable
		${\hat {\bm \sigma}}\cdot {\bm u}^{(\eta)}$ 
		for the copy $\eta$. If the outcome is +1, the experimenter notes down the direction  ${\bm u}^{(\eta)}$ in the notebook. If the outcome is -1, the experimenter records the direction  $-{\bm u}^{(\eta)}$ in the notebook. In the case of this figure the outcome of the measurements 1, 3, 4,... are +1, while those of  2, 5,... are -1, thus the directions ${\bm u}^{(1,3,4,..)}$ and $-{\bm u}^{(2,5,..)}$ are noted down in the notebook.}
	\label{texp2}
\end{figure}

From the above discussion, we know that  the  Husimi function $P_H({\hat \rho}; {\bm n})$ can be interpreted as the conditional probability density ${\cal P}\left({\bm n}\big\vert+1\right)$. Moreover, in the above thought experiment with $m\rightarrow\infty$, we can consider the directions noted down in the notebook as an ensemble of directions in  $S^2$, and  the probability distribution corresponding to this ensemble is just   $P_H({\hat \rho};  {\bm n})$.

Additionally, the Husimi $P_H({\hat \rho}; {\bm n})$ can also be interpreted with another thought experiment, which is  similar to the above one. 
As above, the experimenter  also prepare 
$m$ copies and perform 
$m$ measurements on the observable ${\hat {\bm \sigma}}\cdot {\bm u}^{(\eta)} (\eta=1,...,m)$, and notes down the direction  ${\bm u}^{(\eta)}$ in the notebook if the outcome is $+1$. Nevertheless, if the outcome is $-1$, the experimenter records the direction  $-{\bm u}^{(\eta)}$ in the notebook, rather than just ignore it as in the above though experiment, as shown in  Fig.~\ref{texp2}. With similar calculations as above, we can prove that for $m\rightarrow \infty$ the distribution of the directions in the notebook still satisfies Eq.~(\ref{nn}), {\it i.e.}, is proportional to the Husimi function.

For convenience, our following discussions  are all based  on  the first thought experiment ({\it i.e.}, the one of Fig.~\ref{texp}).

The above discussions can be 
straightforwardly generalized to the multi-particle ($N>1$) cases. 
In the though experiment of these cases,  each copy $c^{(\eta)}$ ($\eta=1,...,m$) includes $N$ particles, and every copy has the same $N$-body density operator ${\hat \rho}$. In addition,
each ${\bm u}^{(\eta)}$ $(\eta=1,...,m)$ is randomly selected from $S^{2\otimes N}$, and includes $N$ component with each one being in $S^2$,
{\it i.e.},  ${\bm u}^{(\eta)}\equiv ({\bm u}_1^{(\eta)},...,{\bm u}_N^{(\eta)})$
 with ${\bm u}_j^{(\eta)}\in S^2$ ($j=1,...,N$). 
As before, the  thought experiment includes  $m$ steps of measurement and noting. Nevertheless, now in the $\eta$-th step ($\eta=1,…,m$), the experimenter needs to measure $N$ observables, {\it i.e.},
measure the value of ${\hat {\bm \sigma}}\cdot{\bm u}^{(\eta)}_j$ 
 of the $j$-th  particle of the copy $c^{(\eta)}$, for all $j=1,...,N$. The experimenter notes down ${\bm u}^{(\eta)}$ in the notebook if and only if the outcomes of these $N$ measurements  are all $+1$.
 
  As in  the single-particle case, here
 we can still prove the relation of Eq.~(\ref{pph}), while now ${\bm n}\equiv({\bm n}_1,...,{\bm n}_N)\in S^{2\otimes N}$, with ${\bm n}_j\in S^2$ ($j=1,...,N$). 
 Therefore, for  arbitrary particle number $N$,
the Husimi function $P_H({\hat \rho};  {\bm n})$ can always  be interpreted as
${\cal P}\left({\bm n}\big\vert +1\right)$, {\it i.e.},
 the probability density of a randomly-chosen element of $S^{2\otimes N}$ being ${\bm n}$, under the condition that the outcomes of the measurements of  ${\hat {\bm \sigma}}\cdot {\bm n}_j$ 
 of the particle $j$  for each particle $j$ ($j=1,...,N$)
are {\it all +1}.  Clearly, it is also the distribution of the  ${\bm u}$-vecotrs noted down in the thought experiment  in the limit $m\rightarrow\infty$.
   
Furthermore, according to the above statistical interpretation  of the Husimi function $P_H({\hat \rho}; {\bm n})$, the Wehrl entropy 
 can be interpreted as the entropy corresponding to the conditional probability density ${\cal P}\left({\bm n}\big\vert +1\right)$ of our system, or the entropy corresponding to the distribution of the  ${\bm u}$-vecotrs noted down in the thought experiment, in the limit that there are  infinity copies.

So far we have obtained the statistical interoperation of the Husimi function and Wehrl entropy of the of $N$ quantum spin-1/2 particles. In Sec.~\ref{cc}, we explore the classical correspondence of the above interpretation of the Wehrl entropy   of spin-1/2 particles.
Precisely speaking, spin-1/2 particles constitute  a pure quantum system and do not have exact  classical correspondences. Nevertheless, as described in Sec.~\ref{cc}, precessing symmetric spinning tops  can be viewed as a classical analogy of spin-1/2 particles. By calculating the related probabilities, we demonstrate that in this analogy, that in this analogy the Wehrl entropy of $N$ quantum spin-1/2 particles corresponds the Gibbs entropy of $N$ such classical tops.

%The results in this section and Appendix~\ref{cc} can be generalized to  high-spin particles as well as systems with continuous variables, {\it e.g.}, the oscillators. These results show not only a statistical explanation of the WE, but also the intrinsic relation between the Wehrl entropy and the  Gibbs entropy in classical physics, and yield that the Wehrl entropy can be considered as a generalization the the latter in quantum physics.

\section{Classical Correspondence of the Husimi Function and Wehrl Entropy}
\label{cc}

In this section we study the classical correspondence of the statistical interpretation of Husimi function and Wehrl entropy  of spin-1/2 particles, which are given in Sec.~\ref{pma}. 

As in that section, we begin from  the single-particle ($N=1$) case. 
Precisely speaking, a spin-1/2 particle is a pure quantum system and does not have exact  classical correspondence. Nevertheless, we can still find  ``classical analogies" for this kind of particle, {\it i.e.}, 
classical systems with dynamic equations being equivalent to the Heisenberg equations of a spin-1/2 particle.

\begin{figure}[tbp]
	\centering
	\includegraphics[width=9cm]{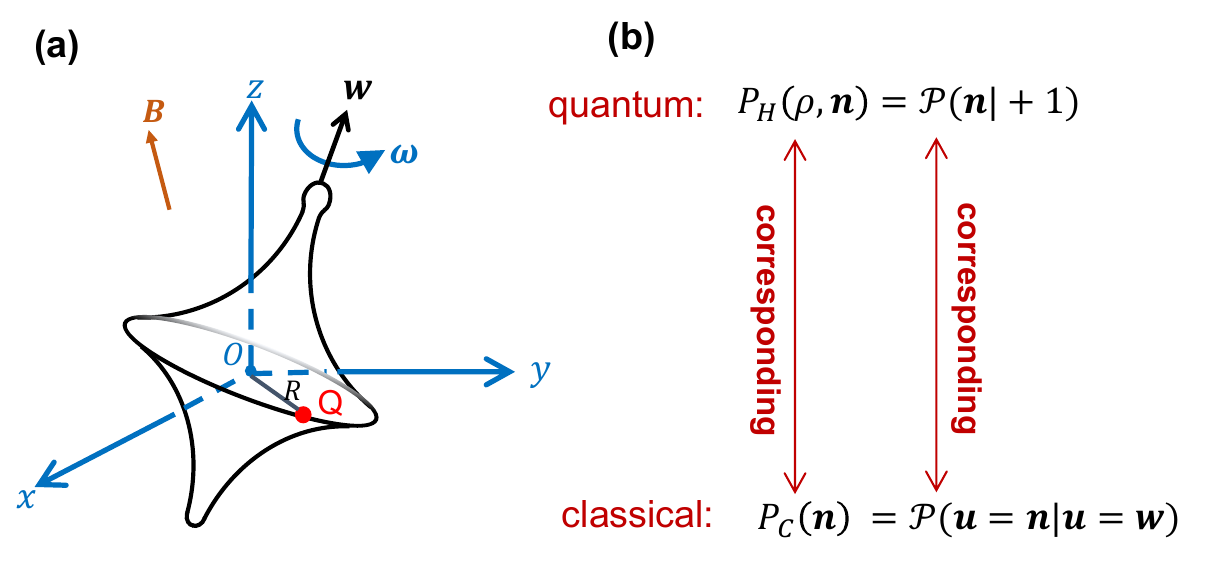}
        \caption{{\bf (a):} A spinning charged top precessing in a homogeneous magnetic field, which is a classical analogy of a spin-1/2 particle. Here $O$ is the center-of-mass of the top, and ${\bm w}$ is the unit vector along the symmetry axis. Other details  are introduced in Secs.~\ref{cc}. {\bf (b):} The identities proven in Secs.~\ref{pma} and Secs.~\ref{cc} for the quantum and classical systems, respectively, and the quantum-classical correspondence of the related functions.}
	\label{top}
\end{figure}

In Fig~\ref{top}(a) we present one example of such systems: a charged axial symmetric top precessing in a homogeneous magnetic field. Explicitly, the top is  rapidly spinning along its symmetry axis with fixed angular speed $\omega$, and the center-of-mass of this top is fixed in an inertial frame of reference. In addition,  an electric charge $Q$ is fixed on the edge of the top, forming an electric current during the spinning. Due to this current, the top gains a magnetic moment, and thus is subject to a force moment from the magnetic field. The precession of this top, {\it i.e.}, the changing of the 
the direction of the symmetry axis of the top in the inertial frame of reference, is caused by this force moment. 
When the spinning is rapid enough, the unit vector ${\bm { w}}$ along the symmetry axis of the top satisfies the dynamical equation during the precession (Appendix~\ref{pr}).
\begin{eqnarray}
\frac{\mathrm{d}{\bm {w}}}{\mathrm{d} t}=C{\bm {w}}\times{\bm B},\label{ctt}
\end{eqnarray}
where ${\bm B}$ is the magnetic field and $C=QR^2/(2I)$. Here
$R$ is the distance between the charge and the symmetric axis of the top, and $I$ is the moment of inertia of the top along this axis.
On the other hand, the Hamiltonian of a quantum spin-1/2 particle can always be written as (up to a constant) a linear combination of Pauli operators $\sigma_{x,y,z}$, {\it i.e.},
$
H_S=\hbar\sum_{\alpha=x,y,z}f_\alpha\sigma_\alpha.
$  Thus, the corresponding Heisenberg equation is 
\begin{eqnarray}
\frac{\mathrm{d}{\hat {\bm {\sigma}}}_H}{\mathrm{d}t}=-2{\hat {\bm {\sigma}}_H}\times{\bm f},\label{sh}
\end{eqnarray}
where ${\hat {\bm \sigma}}_H=({\hat {\sigma}}_{xH}, {\hat {\sigma}}_{yH}, {\hat {\sigma}}_{zH})$ 
is the time-dependent Pauli-operator vector in the Heisenberg picture,
and ${\bm f}=(f_x,f_y,f_z)$. Clearly, this Heisenberg equation is equivalent to the dynamical equation 
 (\ref{ctt}) of the above charged spinning top in a magnetic field ${\bm B}=-2{\bm f}/C$.

 The above analysis reveals that  the  charged spinning top is a classical analogy of a quantum spin-1/2 particle.  The direction vector ${{\bm w}}\equiv(w_x,w_y,w_z)$ of the top axis is just the classical analogy of the Pauli -operator vector ${\hat {\bm \sigma}}\equiv({\hat \sigma}_x,{\hat \sigma}_y,{\hat \sigma}_z)$ of the spin-1/2 particle.
 Furthermore, as shown in Appendix~\ref{ps}, the classical dynamical equation (\ref{ctt}) of the top can be re-expressed as a Hamiltonian equation, and the phase space of this system is just the set of all possible direction vectors of the top axis,  coinciding with the unit sphere 
$S^2$. As a result, the state of this classical top is described by the phase-space probability distribution $P_C({\bm n})$ with ${\bm n}\in S^2$. Explicitly, $P_C({\bm n})$ represents the probability density of the event that the symmetry axis of the top is along ${\bm n}$, {\it i.e.}, ${\bm w}={\bm n}$.

In the discussion in Sec.~\ref{pma}, which is for the spin-1/2 quantum particle, we  introduced the conditional probability density  ${\cal P}\left({\bm n}\big\vert+1\right)$.
Now let us consider the classical correspondence   of ${\cal P}\left({\bm n}\big\vert+1\right)$. As shown above, the classical correspondence of the ${\hat{\bm \sigma}}$-operator is the symmetry axis direction ${\bm w}$. Therefore, for a given direction ${\bm u}\in S^2$, the classical correspondence of the event ${\hat{\bm \sigma}}\cdot {\bm u}=+1$ is the event ${\bm w}={\bm u}$. Therefore,
the classical
correspondence   of ${\cal P}\left({\bm n}\big\vert+1\right)$ is a conditional probability density  ${\cal P}\left({\bm u}={\bm n}\big\vert {\bm w}={\bm u}\right)$, which is defined as
the probability that 
a random direction ${\bm u}$ being in a small solid angle $\Delta\Omega$ around the direction ${\bm n}$, under the condition that ${\bm w}={\bm u}$, is ${\cal P}\left({\bm u}={\bm n}\big\vert {\bm w}={\bm u}\right)\Delta\Omega$. Clearly, ${\cal P}\left({\bm u}={\bm n}\big\vert {\bm w}={\bm u}\right)$  can in principle be measured via a ``classical version" of the thought experiment in Sec.~\ref{pma} \cite{cvtexp}.

Furthermore, for a charged spinning top with phase-space distribution function $P_C({\bm n})$, the conditional probability density  ${\cal P}\left({\bm u}={\bm n}\big\vert {\bm w}={\bm u}\right)$ can be calculated via Bayes' theorem as:
 \begin{eqnarray}
{\cal P}\left({\bm u}={\bm n}\big\vert {\bm w}={\bm u}\right)=\frac{{\cal P}\left({\bm w}={\bm u}\big\vert {\bm u}={\bm n}\right)}{{\cal P}\left({\bm w}={\bm u}\right)}
{\cal P}\left({\bm u}={\bm n}\right),\nonumber\\
\label{bp2}
\end{eqnarray}
where ${\cal P}\left({\bm u}={\bm n}\right)=1/(4\pi)$ is the probability density of the random vector ${\bm u}$ being ${\bm n}$, ${\cal P}\left({\bm w}={\bm u}\big\vert {\bm u}={\bm n}\right)$ is the probability density of ${\bm w}={\bm u}$ under the condition that $ {\bm u}={\bm n}$, and ${\cal P}\left({\bm w}={\bm u}\right)=
\int \mathrm{d}{\bm n}^\prime {\cal P}\left({\bm w}={\bm u}\big\vert {\bm u}={\bm n}^\prime\right){\cal P}({\bm u}={\bm n}^\prime)
$. Because  ${\cal P}\left({\bm u}={\bm n}\right)$ and ${\cal P}\left({\bm w}={\bm u}\right)$ are all independent of ${\bm n}$, Eq.~(\ref{bp2}) yields
 \begin{eqnarray}
{\cal P}\left({\bm u}={\bm n}\big\vert {\bm w}={\bm u}\right)=
\xi{\cal P}\left({\bm w}={\bm u}\big\vert {\bm u}={\bm n}\right),\label{bp3}
\end{eqnarray}
where $\xi$ is the normalization factor determined by the condition $\int d{\bm n} {\cal P}\left({\bm u}={\bm n}\big\vert {\bm w}={\bm u}\right)=1$. Furthermore, we have
\begin{eqnarray}
{\cal P}\left({\bm w}={\bm u}\big\vert {\bm u}={\bm n}\right)&=&
{\cal P}\left({\bm w}={\bm n}\big\vert {\bm u}={\bm n}\right)\nonumber\\
&=&P_C({\bm n}),\label{bp4}
\end{eqnarray}
where the second equality is due to the fact that ${\cal P}\left({\bm w}={\bm n}\big\vert {\bm u}={\bm n}\right)$ is just the probability density of ${\bm w}={\bm n}$, {\it i.e.}, $P_C({\bm n})$. Substituting Eq.~(\ref{bp4}) into Eq.~(\ref{bp3}), and using the normalization condition, we finally obtain
\begin{eqnarray}
{\cal P}\left({\bm u}={\bm n}\big\vert {\bm w}={\bm u}\right)=P_C({\bm n}).\label{eqc}
\end{eqnarray}

%Now we consider the ``classical version" of the thought experiment in Sec.~\ref{pma} for $N=1$, and show the 
%classical correspondence of the Husimi function. As in Sec.~\ref{pma},
%we assume there are
% $m$ copies $t^{(1)}, t^{(2)},...,t^{(m)}$ of the charged spinning top, with each one having the same phase-space probability distribution $P_C({\bm n})$. The experimenter choose $m$ independent random directions ${\bm u}^{(1)}, {\bm u}^{(2)},...,{\bm u}^{(m)}$ from $S^2$, and 
%measure the projection of the symmetry axis direction ${{\bm w}}$ on the direction $ {{\bm u}}^{(\eta)}$ ({\it i.e.},
%the value of ${{\bm w}}\cdot {{\bm u}}^{(\eta)}$) for the  copy $t^{(\eta)}$ ($\eta=1,...,m$).
%If the outcome is +1, then experimenter denote the  direction ${\hat {\bm u}}^{(\eta)}$ in the notebook. Now the question is, after all the measurements, what is the distribution of the denoted directions?
%Notice that we have ${{\bm w}}\cdot {{\bm u}}=1$ if and only if  ${{\bm w}}={{\bm u}}$.
%Using this fact and calculations based on Bayes' theorem, which is similar as the ones in Sec.~\ref{pma}, we find that the distribution of the denoted directions, or the  probability density  ${\cal P}\left({\bm n}\big\vert+1\right)$ of a randomly-chosen direction being ${\bm n}$ under the condition that the outcome of the measurement of ${{\bm w}}$ along that direction is $+1$, is nothing but the phase-probability distribution function $P_C({\bm n})$. 

 As shown above, in our quantum-classical analogy the classical
correspondence   of ${\cal P}\left({\bm n}\big\vert+1\right)$ is a conditional probability density  ${\cal P}\left({\bm u}={\bm n}\big\vert {\bm w}={\bm u}\right)$.
Additionally, as shown in Sec.~\ref{pma}, in the quantum case we have ${\cal P}\left({\bm n}\big\vert+1\right)=P_H({\hat \rho};  {\bm n})$.  Moreover, Eq.~(\ref{eqc}) show that in the classical side we have ${\cal P}\left({\bm u}={\bm n}\big\vert {\bm w}={\bm u}\right)=P_C({\bm n})$. Therefore,  the classical correspondence of the Husimi function $P_H({\hat \rho};  {\bm n})$ is the phase-space probability distribution function $P_C({\bm n})$ (Fig.~\ref{top}(b)).

Using the above conclusion and the relationship  between  the Wehrl entropy  and the Husimi function   ({\it i.e.}, Eq.~(\ref{sw})), we further find that the classical analogy of the Wehrl entropy is $ -\int P_C({\bm n})\ln\left[P_C({\bm n})\right]\mathrm{d}{\bm n}$. This is nothing but the Gibbs entropy of the spinning top because $P_C({\bm n})$ is the phase-space probability distribution function of this top.

The above discussions can be directly generalized to the multi-particle cases with $N>1$. The classical analogy of $N$ spin-1/2 particles are $N$ charged spinning tops, and the $N$-body phase space  is the product $S^{2{\otimes N}}$ of the $N$ spheres. Similar as above, one can find that the classical analogy of the Husimi function and Wehrl entropy of the $N$ quantum particles are the probability distribution of the tops in this $N$-body phase space and the corresponding  Gibbs entropy, respectively.

\section{continuous-variable systems}
\label{cv}

 Above statistical interoperation of  Husimi function and Wehrl entropy of spin systems can be generalized to the systems with  continuous variables. In this section we illustrate this generalization with the example of a 1-dimensional (1D)  particle moving along the $x$-axis. 
 
 The coherent states of this particle can be denoted as $\{|\psi_{x_\ast,p_\ast}\rangle|x_\ast,p_\ast\in\mathbb R\}$. The real-space wave function of the  coherent state $|\psi_{x_\ast,p_\ast}\rangle$ is given by 
 \begin{eqnarray}
	\langle x|\psi_{x_\ast,p_\ast}\rangle=\left(\frac{2}{\pi\sigma^2}\right)^{\frac14}\exp\bigg[-\frac{(x-x_\ast)^2}{\sigma^2}+\frac{\mathrm{i}p_\ast x}{\hbar}\bigg],\nonumber\\
\end{eqnarray}
where $|x\rangle$ is the eigen-state of the coordinate operator. 
 Here $\sigma$ is an $(x_\ast,p_\ast)$-independent positive parameter.  The state $|\psi_{x_\ast,p_\ast}\rangle$ satisfies 
 \begin{eqnarray}
 \langle\psi_{x_\ast,p_\ast}|\hat{x}|\psi_{x_\ast,p_\ast}\rangle&=&x_\ast;\ \ 
  \langle\psi_{x_\ast,p_\ast}|\hat{p}|\psi_{x_\ast,p_\ast}\rangle=p_\ast,
  \\[0.3cm]
  \Delta x&=&\frac \sigma 2;\ \ \ \ \Delta p=\frac{\hbar}{\sigma}=\frac{\frac12\hbar}{\Delta x},
 \end{eqnarray}
 where $\hat x$ and $\hat p$ are the coordinate and momentum operators, respectively, and  $\Delta \zeta\equiv[\langle\psi_{x_\ast,p_\ast}|\hat{\zeta}^2|\psi_{x_\ast,p_\ast}\rangle-\langle\psi_{x_\ast,p_\ast}|\hat{\zeta}|\psi_{x_\ast,p_\ast}\rangle^2]^{1/2}$, ($\zeta=x,p$). These results yields that the wave functions of $|\psi_{x_\ast,p_\ast}\rangle$ in real and momentum spaces are wave packets centered at $x_\ast$ and $p_\ast$, respectively, with widths $ \sigma /2$ and $\hbar/\sigma$, respectively.

The Husimi function with respect to density operator $\hat \rho$ of this particle is 
\begin{eqnarray}
	P_H({\hat \rho};  x_\ast, p_\ast)\equiv\frac{1}{2\pi\hbar}\langle \psi_{x_\ast,p_\ast}|{{\hat \rho}}|\psi_{x_\ast,p_\ast}\rangle.
\end{eqnarray}
It satisfies the normalization condition 
\begin{eqnarray}
	\int \mathrm{d}x_\ast \mathrm{d}p_\ast P_H({\hat \rho};  x_\ast, p_\ast)=1.
	\end{eqnarray}
	The corresponding Wehrl entropy $S_W({\hat \rho})$ is defined as
	\begin{eqnarray}
S_W({\hat \rho})\equiv-\int P_H({\hat \rho};  x_\ast,p_\ast)\ln\bigg[P_H({\hat \rho}; x_\ast,p_\ast)\bigg]
\mathrm{d}x_\ast \mathrm{d}p_\ast. \nonumber\\
\label{swc}
\end{eqnarray}

%%%%%%%%%%%%%%%%%%%%%%%%%%%%%%%%%%%%%%%%%%%%%%%%%%%%%%%%%%%%%%%%%%%%%%%%%%%

We can obtain a statistical interoperation  of $P_H({\hat \rho})$ via generalizing the thought experiment of Sec.~\ref{cc} to the current
 continuous-variable system. Firstly, we assume  there are $m$ copies $c^{(1)}, c^{(2)},...,c^{(m)}$ of the particle, with each one being in the same state ${\hat \rho}$. Secondly, the experimenter generates $m$ independent random points $(x_\ast^{(1)},p_\ast^{(1)}), (x_\ast^{(2)},p_\ast^{(2)}),...,(x_\ast^{(m)},p_\ast^{(m)})$ in the phase space.  After that, the experimenter performs the following $m$ steps:
In the $\eta$-th  step ($\eta=1,...,m$), the experimenter  measures whether the particle is in the coherent state corresponding to $(x_\ast^{(\eta)},p_\ast^{(\eta)})$, {\it i.e.}, measure
the operator $|\psi_{x_\ast^{(\eta)},p_\ast^{(\eta)}}\rangle\langle\psi_{x_\ast^{(\eta)},p_\ast^{(\eta)}}|$ for the  copy $c^{(\eta)}$. The outcome of this measurement is either 0 and 1, since this operator has only these two eigenvalues.
If and only if the outcome of this  measurement  is $1$,   the experimenter notes down the point $(x_\ast^{(\eta)},p_\ast^{(\eta)})$ in a notebook. 
%In Fig.~xxx we schematically  illustrate these $m$ steps.

We   consider the cases where $m$ is very large.  After all these m steps are finished, we can ask: what is the distribution of the  points noted down in the notebook? As in Sec.~\ref{cc}, using the Bayes' theorem one can directly prove that 
the density distribution of  noted points in the phase space  is just (up to a normalization constant) the Husimi function $P_H({\hat \rho};  x_\ast, p_\ast)$ with respect to the density operator ${\hat \rho}$ of the particle.  Consequently, the corresponding Wehrl entropy $S_W({\hat \rho})$ just describes the degree of complexity of this density distribution.

Now we investigate the classical correspondence of the Husimi function and Wehrl entropy. To this end, as in 
Sec.~\ref{cc}, we study the classical analogy of the above thought experiment. 
We consider a 1D classical particle with coordinate $x$ and momentum $p$, respectively. The phase-space probability distribution ({\it i.e.}, the Liouville function) of this particle is $\rho_L(x_\ast, p_\ast)$. In the though experiment, the experimenter prepares $m$ copies $c^{(1)}, c^{(2)},...,c^{(m)}$ of this particle,
and generates $m$ independent random point $(x_\ast^{(1)},p_\ast^{(1)}), (x_\ast^{(2)},p_\ast^{(2)}),...,(x_\ast^{(m)},p_\ast^{(m)})$ in the phase space.  After that, the experimenter performs the following $m$ steps:
In the $\eta$-th  step ($\eta=1,...,m$), the experimenter  measures the position and momentum of the  copy $c^{(\eta)}$.
If and only if the outcome of this  measurement is $x=x_\ast^{(\eta)}$ and $p=p_\ast^{(\eta)}$, the experimenter notes down the point $(x_\ast^{(\eta)},p_\ast^{(\eta)})$ in a notebook.

We   consider the case that the steps $m$ is very large.  After all these steps are completed, the distribution of the points noted down in the notebook is the classical correspondence of the Husimi function $P_H$.
On the other hand, with the Bayes' theorem it can be proven that 
this density distribution is nothing but the Liouville function $\rho_L(x_\ast, p_\ast)$. Therefore, the classical 
correspondence of the Husimi function $P_H$ is the Liouville function $\rho(x_\ast, p_\ast)$. Accordingly, the  classical 
correspondence of the Wehrl entropy $S_W$ is the Gibbs entropy, which is defined as
$S_G= -\int \mathrm{d}x_\ast \mathrm{d} p_\ast \rho_L(x_\ast, p_\ast)\ln\left[\rho_L(x_\ast, p_\ast)\right]$ and plays important role in statistical physics. 

%The limitation of this interpretation lies in the fact that the definition of coherent states is not unique. Therefore, both the definitions of the Husimi function and Wehrl entropy are not unique and depend on the definition of coherent states, which is an issue that needs to be considered in applications. Obviously, this issue does not exist in the classical limit when $\hbar\to0$, where the region with area {\color{red}$\hbar/2$} in the quantum case will shrink to a point in the classical limit.

\section{Summary}
\label{summary}
In this work  we provide a  statistical interpretation to Husimi function and Wehrl entropy for spin-1/2 particles as well as systems with continuous variables, which is based on  Bayes' theorem. With this interpretation, we show that the classical correspondence of Husimi function and Wehrl entropy are just the Liouville function and Gibbs entropy, respectively. 
This function and entropy are being increasingly utilized in the studies of various quantum systems, and 
our results  are helpful for understanding the properties of these systems.

\begin{acknowledgments}
	The authors thank Yingfei Gu, Dazhi Xu, Pengfei Zhang, Hui Zhai, Honggang Luo, Zhiyuan Xie, and Ninghua Tong  for fruitful discussions.
	This work was supported by the  National Key Research and Development Program of China [Grants No. 2022YFA1405300 (P.Z.)], and Innovation Program for Quantum Science and Technology (Grant No.~2023ZD0300700).
\end{acknowledgments}

 \begin{figure}[h]
	\centering
	\includegraphics[width=5.5cm]{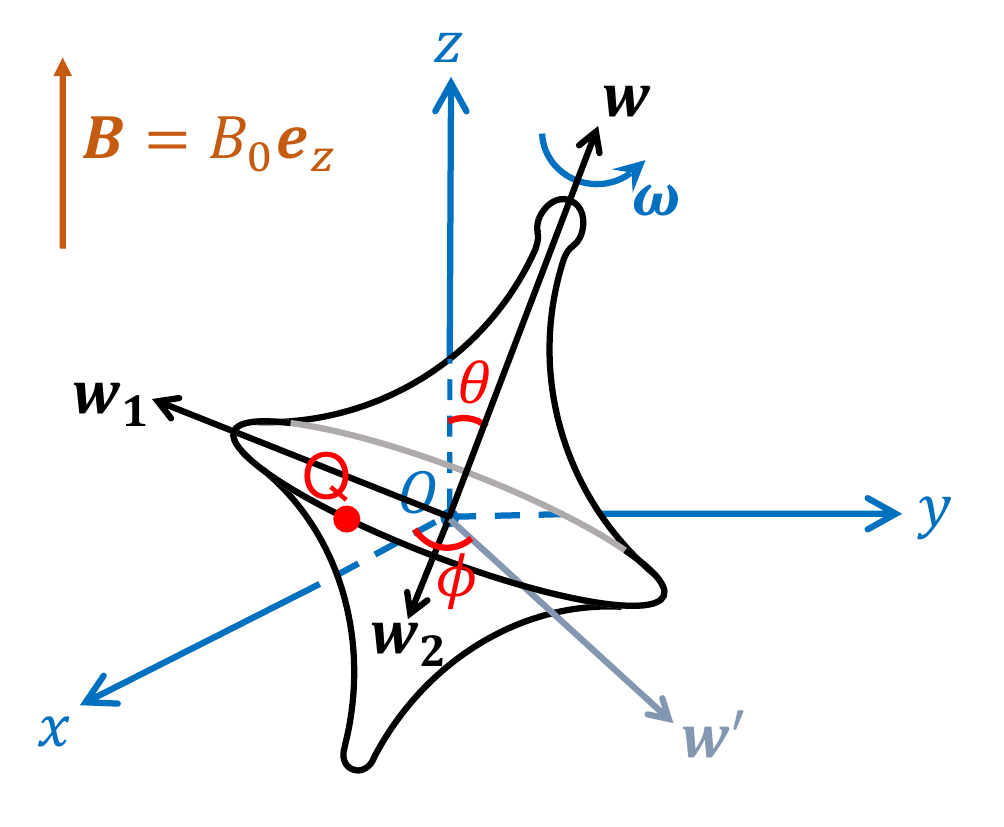}
	\caption{A spinning charged top precessing in a homogeneous magnetic field $\bm B=B_0\bm{e}_{z}$. Here $O$ is the CoM of the top, $O$-$xyz$ is the CoM frame, ${\bm w}$ is the unit vector along the symmetry axis of the top. The unit vectors $\bm{w}_1$ and $\bm{w}_2$ are mutually perpendicular unit vectors, both of which are also perpendicular to ${\bm w}$. Other details  are introduced in Appendix~\ref{pr}.}
	\label{top2}
\end{figure}

\appendix

\section{Proof of Eq.~(\ref{ctt})}
\label{pr}
In this appendix, we prove that when the spinning of the top is rapid enough, the unit vector ${\bm { w}}$ along the symmetry axis of the top satisfies the dynamical equation (\ref{ctt}) during the precession.

We first clarify the specific meaning of the above condition ``spinning of the top is rapid enough". As shown in Fig.~\ref{top2}, we denote $O$-$xyz$ as the center-of-mass (CoM) frame of the top.  Without loss of generality, we assume $\bm B$ is along $z$-direction, {\it i.e.}, $\bm B=B_0\bm{e}_{z}$.
The angles $\theta$ and $\phi$ and polar and azimuthal angles of the direction ${\bm w}$ of the symmetric axis, respectively. During the motion, the top is spinning around ${\bm w}$. In addition, both $\phi$ and $\theta$ can also change with time, {\it i.e.}, there can be precession and nutation. The aforementioned ``spinning is rapid enough" means that the following two conditions are satisfied simultanesouly: 
\begin{itemize}
\item  [(i)] $|\omega|\gg |\dot \phi|, |\dot \theta|$.

\item [(ii)] The contribution of the precession and nutation to the angular ${\bm L}$ of the top in the CoM frame is negligible. Namely, ${\bm L}$ can be approximated as 
\begin{eqnarray}
\bm L=\omega I{\bm w},\label{lapp}
\end{eqnarray}
where $I$ is the corresponding moment of inertia.

\end{itemize}
 
 Now we consider the dynamics of the top. 
According to the angular momentum theorem in the CoM frame, we have
\begin{eqnarray}
	\frac{\mathrm{d} \bm{L}}{\mathrm{d} t}=\bm{M},\label{dL0}
\end{eqnarray}
where $\bm M$ is 
the instantaneous  torque of the Lorentz force on the charge $Q$ in the CoM frame. Due to the above condition (ii), we have
%\begin{eqnarray}
%\frac{d\bm{L}}{dt}=I\dot\omega\bm w+I\omega\frac{d\bm w}{dt}.\label{dL}
%\end{eqnarray}
\begin{eqnarray}
	\frac{\mathrm{d} \bm{L}}{\mathrm{d} t} = I \dot{\omega} \bm{w} + I \omega \frac{\mathrm{d} \bm{w}}{\mathrm{d} t}.\label{dL}
\end{eqnarray}
Furthermore, due to the above condition (i), we can approximately replace the instantaneous torque ${\bm M}$ in Eq.~(\ref{dL0}) with the time-averaged value  over a period $T=2\pi/\omega$:
\begin{eqnarray}
	\langle\bm M\rangle=\frac1 T\int_0^T\mathrm{d} t\bm M(t).
\end{eqnarray}
Using Eq.~(\ref{dL}) and this approximation, we have
\begin{eqnarray}
I\dot\omega\bm w+I\omega\frac{\mathrm{d}\bm w}{\mathrm{d} t}=\langle\bm M\rangle.\label{dL3}
\end{eqnarray}

Now we calculate the average torque $\langle\bm M\rangle$. 
To this end, we first notice that the direction vector $\bm w$ can be expressed as
\begin{eqnarray}
	\bm w=\sin\theta\cos\phi{\bf e}_{x}+\sin\theta\sin\phi{\bf e}_{y}+\cos\theta{\bf e}_{z},
\end{eqnarray}
with ${\bf e}_{x,y,z}$ being the unit vectors along the $x,y,z$ directions of the CoM frame, respectively. We further introduce unit vectors  $\bm{w}_1$, $\bm{w}_2$
\begin{eqnarray}
	\bm w_1&=&-\sin\phi\bm{e}_{x}+\cos\phi\bm{e}_{y},\\
	\bm w_2&=&-\cos\theta\cos\phi\bm{e}_{x}-\cos\theta\sin\phi\bm{e}_{y}+\sin\theta\bm{e}_{z},
\end{eqnarray}
which are both perpendicular to $\bm w$, and are perpendicular with each other. 
The instantaneous torque ${\bm M}(t)$ is given by 
\begin{eqnarray}
\bm M(t)=\bm F_L(t)\times\bm r(t),
\end{eqnarray}
where $\bm r(t)$ and $\bm F_L(t)$ are the charge position and the Lorentz force, respectively. Explicitly, we have
\begin{eqnarray}
	\bm r(t)=R\cos(\omega t)\bm{w}_1+R\sin(\omega t) \bm{w}_2,
\end{eqnarray}
and 
\begin{eqnarray}
\bm F_L(t)=Q \bm v(t)\times \bm B,
\end{eqnarray}
where
\begin{eqnarray}
\bm v(t)=\frac{\mathrm{d}}{\mathrm{d} t}\bm r(t).
\end{eqnarray}
Using these relations and straightforward calculations, we obtain
\begin{eqnarray}
	\langle\bm M\rangle=\frac12QR^2\omega\bm w\times\bm B.\label{meanM}
\end{eqnarray}

Substituting Eq.~(\ref{meanM}) into Eq.~(\ref{dL3}), and using the fact that $d\bm w/dt\perp \bm w$, which is due to the fact that the norm of $\bm w$ does not change with time,
we directly obtain Eq.~(\ref{ctt}) of our main text and $\dot \omega=0$.

In the end of this appendix, we point out that the mechanism of the top dynamics shown above is very similar to the one of a rapidly spinning electrically neutral top processing the gravity \cite{TM}.

%[cite: S. Lu, J. Hu and J. Guan, {\it Theoretical Mechanics}, Publishing House of Electronics Industry, 1991 (Chinese).]

%%%%%%%%%%%%%%%%%%%%%%%%%%%%%%%%%%%%%%%%%%%%%%%%%%%%%%%%%%%%%%%%%%%%%%%%

\section{The Phase Space of  Charged Spinning Top}
\label{ps}

In this appendix  we show that the phase space of the charged spinning top studied in  Sec.~\ref{cc} is just the unit sphere $S^2$.

The canonical coordinate of this top can be chosen as the azimuth angle $\varphi$ and the $z$-component $w_z$ of the direction vector ${\hat{\bm w}}$ of the top axis (Fig.~\ref{top}(a)). Then the Lagrangian of this system is given by 
\begin{eqnarray}
L_{\rm top}\!\!&=&\!\!C\left[B_zw_z\!+\!B_x\sqrt{1-w_z^2}\cos\varphi+\!B_y\sqrt{1-w_z^2}\sin\varphi\right]\nonumber\\
&&\!\!+\dot{\varphi}w_z.
\end{eqnarray}
It can be straightforwardly proven that the Lagrangian equations
\begin{eqnarray}
\frac {\mathrm{d}}{\mathrm{d}t}\left(\frac{\partial L_{\rm top}}{\partial \dot{\varphi}}\right)-\frac{\partial L_{\rm top}}{\partial {\varphi}}&=&0;\label{l1}\\
\frac {\mathrm{d}}{\mathrm{d}t}\left(\frac{\partial L_{\rm top}}{\partial \dot{w_z}}\right)-\frac{\partial L_{\rm top}}{\partial {w_z}}&=&0,\label{l2}
\end{eqnarray}
are equivalent to the classical dynamical equation (\ref{ctt}) of this top, and the $x$ and $y$ components $w_{x,y}$ of the direction vector ${\hat{\bm w}}$ are functions of the canonical coordinates, {\it i.e.}, $w_x=\sqrt{1-w_z^2}\cos\varphi$ and $w_y=\sqrt{1-w_z^2}\sin\varphi$.

Moreover, the canonical momentums $p_{\varphi}$ and $p_w$  of our system are canonical coordinates
\begin{eqnarray}
p_{\varphi}=\frac{\partial L_{\rm top}}{\partial \dot{\varphi}}&=&w_z;\\
p_{w_z}=\frac{\partial L_{\rm top}}{\partial \dot{w_z}}&=&0.
\end{eqnarray}
Therefore, although there are two canonical coordinates, the phase space of our system is
two-dimensional, rather than four-dimensional. Explicitly, this phase space is
the space of $(\varphi,w_z)$, with $\varphi\in[0,2\pi)$ and $ w_z\in[-1,1]$. In addition, the measurement (area element) of this phase space is $d\varphi dw_z$. It is clear that this phase space is equivalent to the unit sphere $S^2$ with azimuth angle being $\varphi$ and polar angle being $\theta\equiv\arccos w_z=\arccos p_\varphi$,
{\it i.e.}, the set of all possible direction vectors ${\hat{\bm w}}$.
 Moreover, the measurement $d\varphi dw_z$ just equals to $\sin\theta d\theta d\varphi$, {\it i.e.}, the area element of the $S^2$ sphere.

Additionally, Hamiltonian of this charged spinning top:
\begin{eqnarray}
&&H_{\rm top}\nonumber\\
&=&-C\left[B_zp_{\varphi}+B_x\sqrt{1-p_{\varphi}^2}\cos\varphi+B_y\sqrt{1-p_{\varphi}^2}\sin\varphi\right].\nonumber\\
\end{eqnarray}
The Hamilton equation with respect to $H_{\rm top}$ is equivalent to the the Lagrangian equations (\ref{l1}, \ref{l2}), and thus equivalent to  to the classical dynamical equation
(\ref{ctt}).

%\begin{widetext}
\bibliography{reference.bib} % code #2

\end{document}